# The LHeC Project: e-Ring Revisited

Umit KAYA[1,2], Bora KETENOGLU[*,2], Saleh SULTANSOY[1,3]

[1]*TOBB University of Economics and Technology, 06560, Ankara, Turkey*
[2]*Ankara University, 06100, Ankara, Turkey*
[3]*ANAS Institute of Physics, Baku, Azerbaijan*
*\*corresponding author e-mail: bketen@eng.ankara.edu.tr*

**Abstract:** Construction of a new 9 km long e-ring tangential to the Large Hadron Collider (LHC) has been proposed as an option for QCD-Explorer stage of the Large Hadron electron Collider (LHeC). It is shown that $L=10^{33}$ cm$^{-2}$s$^{-1}$ can be achieved with 90 MW synchrotron radiation losses. This luminosity value, which coincides with basic version of ERL60⊗LHC, will be sufficient for precise determination of (Parton Distribution Function) PDFs for LHC, as well as exploration of QCD basics, especially small *x* Björken region up to $10^{-6}$ at $Q^2 \approx 1$ GeV$^2$. In addition, some comments on basic and upgraded versions of ERL60⊗LHC are presented as well. It is shown that upgraded ERL60⊗LHC version with $L=10^{34}$ cm$^{-2}$s$^{-1}$ requires high wall plug power exceeding 160 MW.

*Key words*: Ep collider, Luminosity, LHeC project, Electron-hadron scattering

## LHeC Projesi: e-Halka'nın Yeniden Gözden Geçirilmesi

**Özet:** Büyük Hadron elektron Çarpıştırıcısı (LHeC)'in QCD-Explorer aşaması için bir seçenek olarak, Büyük Hadron Çarpıştırıcısı (LHC)'ye teğet olacak şekilde 9 km uzunluğunda yeni bir e-halkası inşası önerilmiştir. 90 MW sinkrotron ışınımı kaybı ile $L=10^{33}$ cm$^{-2}$s$^{-1}$ ışınlığa ulaşılabileceği gösterilmiştir. ERL60⊗LHC'nin temel versiyonu ile örtüşen bu ışınlık değeri, LHC için (Parton Dağılım Fonksiyonu) PDF'lerin kesin olarak tayini, QCD temellerinin keşfi, ve özellikle $Q^2 \approx 1$ GeV$^2$'de $10^{-6}$'ya kadar küçük *x* Björken bölgesinin belirlenmesinde yeterli olacaktır. Ayrıca, ERL60⊗LHC'nin temel ve geliştirilmiş versiyonları üzerine bazı öneriler de sunulmuştur. Geliştirilmiş ERL60⊗LHC'nin $L=10^{34}$ cm$^{-2}$s$^{-1}$ ışınlıklı versiyonu, 160 MW'ın üzerinde yüksek bir şehir şebeke elektrik gücü gerektirdiği gösterilmiştir.

*Anahtar kelimeler*: Ep çarpıştırıcısı, Işınlık, LHeC projesi, Elektron-hadron saçılması

## 1. Introduction

Design studies for the Large Hadron electron Collider (LHeC) project have been carried out since 2007, under auspices of the European Committee for Future Accelerators (ECFA). The project is planned for collisions of 7 TeV LHC protons with 60-140 GeV electrons/positrons. A luminosity range of $10^{32}$-$10^{33}$ cm$^{-2}$s$^{-1}$ is aimed for various physics goals. Concerning its collision scheme, three options were considered for realization of the LHeC collider [1]: a ring-ring (RR) collider with a new lepton ring in the existing LHC tunnel, a linac-ring (LR) collider based on single pass linac (SPL) and a LR collider based on superconducting energy recovery linac (ERL). Actually all three options give opportunity to achieve luminosities of order of $10^{33}$ cm$^{-2}$s$^{-1}$ with less than 100 MW wall-plug power [2]. However, ERL option is considered as a sole one today [3].

Concerning historical evolution of LHC-based ep colliders, while LEP⊗LHC [4] was considered in 1980s, Linear Collider LC⊗LHC and/or Single Pass Linac SPL⊗LHC options (see review [5] and references therein) have been in the news since 1990s. Afterwards, "LEP"⊗LHC was resurrected at the beginning of 2000s [6]. During 2010s, 60 GeV ERL⊗LHC option has been under consideration [1]. Finally, LC⊗LHC seems to come back in 2020s [2].

As mentioned in [7], QCD-Explorer stage of the LHeC should have high(est) priority for two reasons:
1) HERA provided PDFs for Tevatron and LHC. In the same manner QCD-Explorer will provide PDFs for HL-LHC, HE-LHC and FCC/SppC.
2) Clarify the nature of strong interactions from parton to nuclear level and, consequently, opportunity to enlighten the origin of the 98.5% portion of the visible Universe's mass.

In this paper, we propose construction of a new 60 GeV e-ring with 9 km circumference, like ERL60 option, tangential to LHC. In Section 2, we give some comments on basic and upgraded versions of ERL60⊗LHC. Main parameters of the proposed collider are presented in Section 3. Finally, in Section 4 we present our conclusions and recommendations.

## 2. Comments on LHeC's ERL60⊗LHC Option

Collider parameters of this option given in LHeC CDR [1] (see also [3]) are presented in Table 1. With this parameter set, L=$10^{33}$ cm$^{-2}$s$^{-1}$ and $\sqrt{s}$=1.3 TeV are obtained. Luminosity expression for transversely matched electron and proton beams is given by:

$$L_{ep} = \frac{1}{4\pi e} \frac{N_p}{\varepsilon_p} \frac{1}{\beta_p^*} I_e H_{hg} H_D \qquad (1)$$

where $e$ denotes the electron charge, $N_p$ the proton bunch population, $\beta_p^*$ the proton IP beta function, $I_e$ the electron beam current, $H_{hg}$ (~0.9) the geometric loss factor arising from hourglass effect and $H_D$ (~1.3) the disruption enhancement factor due to the electron pinch in collision.

**Table 1.** Collider Parameters of LHeC's ERL60⊗LHC option with L=$10^{33}$ cm$^{-2}$s$^{-1}$

| Parameter [Unit] | Protons | Electrons |
|---|---|---|
| Beam energy [GeV] | 7000 | 60 |
| Normalized emittance, $\gamma\varepsilon_{x,y}$ [µm] | 3.75 | 50 |
| Beta function @ IP, $\beta^*_{x,y}$ [m] | 0.1 | 0.12 |
| RMS beam sizes @ IP, $\sigma^*_{x,y}$ [µm] | 7 | 7 |
| Bunch length, $\sigma_z$ [mm] | 75 | 0.3 |
| Beam current [mA] | 860 | 6.4 |
| Bunch spacing [ns] | 25 | 25 |
| Bunch population | $1.7 \times 10^{11}$ | $1 \times 10^9$ |

Electrical power consumption estimation is given in Table 2 (Table 7.2 in [1]), where a number of misprints are corrected. As a result, total power consumption is 88.3 MW instead of 75.3 MW (value given in Table 7.2 of the LHeC CDR).

**Table 2.** ERL Power Budget [1]

| Parameter | Electrical Power [MW] | |
|---|---|---|
| | Correct Values | Table 7.2 of LHeC CDR |
| Main linac cryopower | 28.9 | 18.0 |
| Microphonics control | 22.2 | 22.2 |
| Extra RF to compensate SR losses | 24.1 | 24.1 |
| Extra RF cryopower | 1.6 | 1.6 |
| Compensating RF cryopower | 2.1 | - |
| Electron injector | 6.4 | 6.4 |
| Arc magnets | 3.0 | 3.0 |
| Total | 88.3 | 75.3 |

Following the CDR, some modifications of ERL60⊗LHC for higher luminosities are proposed. For instance, upgraded parameters to achieve $10^{34}$ cm$^{-2}$s$^{-1}$ luminosity are given in Table 3 [3]. As clearly seen, number of electrons per bunch is multiplied by factor 4. Considering the power consumption issues in Table 2, the SR losses will increase by a factor 4 (96.4 MW instead of 24.1 MW) and for this case the total power consumption attains more than 160.6 MW value. However, wall plug power was decided to be less than 100 MW for all options [1].

**Table 3.** Collider Parameters of LHeC's ERL60⊗LHC option with L=$10^{34}$ cm$^{-2}$s$^{-1}$

| Parameter [Unit] | Protons | Electrons |
|---|---|---|
| Beam energy [GeV] | 7000 | 60 |
| Normalized emittance, $\gamma\varepsilon_{x,y}$ [µm] | 2.5 | 20 |
| Beta function @ IP, $\beta^*_{x,y}$ [m] | 0.05 | 0.1 |
| RMS beam sizes @ IP, $\sigma^*_{x,y}$ [µm] | 4 | 4 |
| Bunch length, $\sigma_z$ [mm] | 75 | 10 |
| Beam current [mA] | 1112 | 25 |
| Bunch spacing [ns] | 25 | 25 |
| Bunch population | $2.2 \times 10^{11}$ | $4 \times 10^9$ |

### 3. e-Ring Revisited

Main parameters of LHeC CDR's RR option are summarized in Table 4 (Table 6.33 in Ref. 1). Number of electron and proton bunches are equal to 2808 (bunch spacing is 25 ns). Since it is not possible to construct both electron and proton rings in the same LHC tunnel, this option has been forsaken. Actually this situation was obvious from the beginning: LEP⊗LHC was abandoned by the same reason. Two interaction region options in Table 4 correspond to two detector designs: first option covers polar angle acceptance to about 1º (179º) in the forward (backward) direction, second option covers 10º – 170º polar angle region.

**Table 4.** Parameters of LHeC RR option [Table 6.33 of Ref. 1]

| IR Option | 1 Degree | | 10 Degrees | |
|---|---|---|---|---|
| Beams | Electrons | Protons | Electrons | Protons |
| Energy [GeV] | 60 | 7000 | 60 | 7000 |
| Bunch population | $2\times10^{10}$ | $1.7\times10^{11}$ | $2\times10^{10}$ | $1.7\times10^{11}$ |
| Beam current [mA] | 100 | 860 | 100 | 860 |
| $\beta_x^*$ [m] | 0.4 | 4.0 | 0.18 | 1.8 |
| $\beta_y^*$ [m] | 0.2 | 1.0 | 0.1 | 0.5 |
| $\varepsilon_x$ [nm] | 5 | 0.5 | 5 | 0.5 |
| $\varepsilon_y$ [nm] | 2.5 | 0.5 | 2.5 | 0.5 |
| $\sigma_x$ [μm] | 45 | | 30 | |
| $\sigma_y$ [μm] | 22 | | 15.8 | |
| Crossing angle [mrad] | 1 | | 1 | |
| Luminosity [$cm^{-2} s^{-1}$] | $7.33 \times 10^{32}$ | | $1.34 \times 10^{33}$ | |

Here, we propose construction of a new 9 km long ring, which equals to the total length of ERL60, tangential to the LHC (see Fig. 1). The structure of the electron beam is exactly the same as in Table 4. Only difference is that the number of bunches has to be 3 times reduced due to the tunnel length is reduced by a factor of 3.

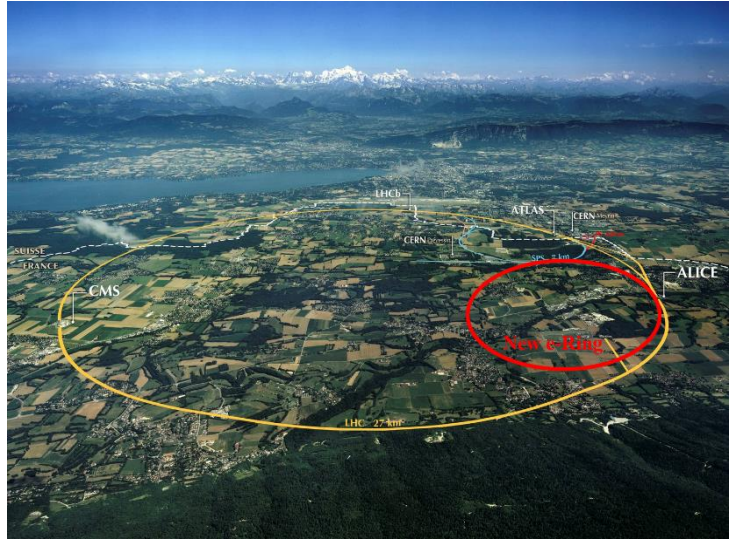

**Fig. 1.** Bird's eye view of proposed ring-ring option for LHeC

When it comes to synchrotron radiation power losses for circular accelerators, the power radiated by a beam of average current *I* is given by:

$$P_{electron}[kW] = 88.46 \frac{E^4[GeV]^4 I[A]}{\rho[m]} \quad (2)$$

where $\rho$ is the bending radius. For e-ring in the LHC tunnel, $\rho$ = 2420 m, *I* = 100 mA and consequently *P* = 47.3 MW. For new 9 km long tunnel, we assume $\rho$ = 1270 m (1 km length is reserved for interaction region, injection and beam dump straight sections). Therefore, *P* = 90 MW.

It should be mentioned that interest in the LHeC project has groving steadily. One of the main reasons for this is that LHeC will provide precision parton distribution functions for HL-LHC [8], HE-LHC [9] as well as future hadron colliders FCC [10] and SppC [11]. On the other hand, activities on physics search potential (see for example [12-18]) of the LHeC have provided new arguments in favor of its construction as well.

## 4. Conclusion and Comments

As mentioned in Introduction, QCD-Explorer stage of the LHeC should have high(est) priority. It should be emphasized that L=$10^{33}$ cm$^{-2}$s$^{-1}$ is sufficient for precise determination of PDFs as well as exploration of QCD basics, especially small *x* Björken region up to $10^{-6}$ at $Q^2 \approx 1$ GeV$^2$. On the other hand, modified ERL60⊗LHC with L=$10^{34}$ cm$^{-2}$s$^{-1}$ requires too much wall plug power, while even this luminosity may not be sufficient for precision Higgs boson physics. In this respect, construction of an additional 9 km e-ring should be considered as a serious alternative for QCD-Explorer stage of the LHeC. An essential advantage of this option is that it is based on well-known technology.

Umit KAYA, umit.kaya@cern.ch, ORCID: https://orcid.org/0000-0003-0823-3848
Bora KETENOGLU, bketen@eng.ankara.edu.tr, ORCID: https://orcid.org/0000-0003-0910-0473
Saleh SULTANSOY, ssultansoy@etu.edu.tr, ORCID: https://orcid.org/0000-0003-2340-748X